\documentclass[aps,pra,reprint,superscriptaddress,showpacs]{revtex4-1}

\usepackage{graphicx}
\usepackage{amsmath}
\usepackage{bm}
\usepackage{graphicx}
\usepackage{dcolumn}
\usepackage{bm}
\usepackage{colordvi}
\usepackage{color}
\usepackage{ulem}
\usepackage{type1cm}
 \usepackage{lettrine}
\newcommand{\beq}{\begin{equation}}
\newcommand{\eeq}{\end{equation}}
\newcommand{\beqa}{\begin{eqnarray}}
\newcommand{\eeqa}{\end{eqnarray}}

\begin{document}

\title{Graphene-assisted resonant transmission and enhanced Goos-H\"{a}nchen shift in a frustrated total internal reflection configuration}

\author{Yi Chen}
\affiliation{Department of Physics, Shanghai University, 200444 Shanghai, People's Republic of China}

\author{Yue Ban}
\affiliation{School of Material, Science and Engineer, Shanghai University, 200444 Shanghai, People's Republic of China}

\author{Qi-Biao Zhu}
\affiliation{Department of Physics, Shanghai University, 200444 Shanghai, People's Republic of China}

\author{Xi Chen}
\email{xchen@shu.edu.cn}
\affiliation{Department of Physics, Shanghai University, 200444 Shanghai, People's Republic of China}

\date{\today}
\begin{abstract}
Graphene-assisted resonant transmission and enhanced Goos-H\"{a}nchen shift are investigated in a two-prism frustrated-total-internal-reflection configuration. Due to the excitation of surface plasmons induced by graphene in low terahertz frequency range, there exist the resonant transmission and anomalous Goos-H\"{a}nchen shifts in such optical tunneling configuration. As compared to the case of quantum well, graphene sheet with unique optical properties can enhance the resonant transmission with relatively low loss, and modulate the large negative and positive Goos-H\"{a}nchen shifts by adjusting chemical potential or electron relaxation time. These intriguing phenomena may lead to some potential applications in graphene-based electro-optic devices.

\end{abstract}
%

\maketitle

In modern optics, frustrated total internal reflection (FTIR) has been investigated theoretically and experimentally for many years, with the applications in optical devices and integrated optics,
including laser resonators, transmission output coupler, and optical filters \cite{AJPReview}. As a matter of fact, FTIR is analogous to the quantum mechanical tunneling, therefore the tunneling time \cite{Steinberg-C,Balcou,Lee,Stahlhofen,Hooper,Gehring} and Goos-H\"{a}nchen (GH) shift \cite{Hsue-T,Riesz-S,Haibel,Winful-Zhang,ChenPRA09,ChenOL} in such two-prism configuration have attracted a lot of attention. Due to the feature of quantum tunneling, the GH shift is significantly reduced in FTIR by coupling of the light into the second prism in the symmetric two-prism configuration with low transmission amplitude. The smallness of GH shift impedes its experimental observation and even application, especially in the optical domain. So the enhancement of transmission amplitude and
GH shift is demanding for the practical use. To this end, several proposals have been put forward
to enlarge the GH shifts by coating quantum wells (metal) \cite{Broe} or dielectric thin film \cite{CFLJAP} onto the surface of two prisms. As compared to the usual case of FTIR,
the GH shift can be significantly enhanced, by transmission resonance, up to more than ten times order of wavelength, which could be useful to design optical sensors and optical switches.

Recently, graphene, a single layer of carbon atoms arranged in a hexagonal lattice \cite{Science,Neto,Peres-Rev}, provides a revolutionary optical material with fantastic optical properties for terahertz to mid-infrared applications in photonics and optoelectronics,
see recent review \cite{Bao,Tony}. In graphene, the charge carrier density and optical conductivity can be feasibly controlled by the bias voltage, resulting in the outstanding optical properties such as strong light-graphene interaction, broadband and high-speed operation etc. \cite{Liu,Bonaccorso}. The unique optical properties have been studied in the spectrum ranging from far-infrared \cite{falkovsky} to visible \cite{stauber} part of the conductivity spectrum. And the electromagnetic scattering problem further shows the universal absorbance of monolayer graphene, and the reflectance and transmittance of graphene are determined by the fine-structure constant. With the potential applications in graphene-based electro-optic devices, the tunable GH shifts and even Imbert-Fedorov shift are extensively studied with modification in various configurations, i.e. at a single interface between dielectric media and graphene \cite{JiangIEEE,XLiOL,GroschearXiv,Hermosa}, attenuated total reflection \cite{MartinezEPL,Cheng}, and photonic crystal with defect \cite{MadaniSuperlattices}.

In this Letter, we shall investigate the resonant transmission and GH shift in FITR configuration coated by the graphene sheet. The GH shift can be significantly enhanced by graphene sheet, due to transmission resonance relevant to surface plasmon excitation in low terahertz (THz) frequency domain. Thanks to the tunable optical properties of graphene, the GH shift can be controlled by adjusting chemical potential and electron relaxation time. Furthermore, comparison is also made with the similar FITR configuration coated by metal quantum well, showing the advantage of graphene. 

\begin{figure}[tbp]
\centering
\fbox{\includegraphics[width=\linewidth,height=8cm]{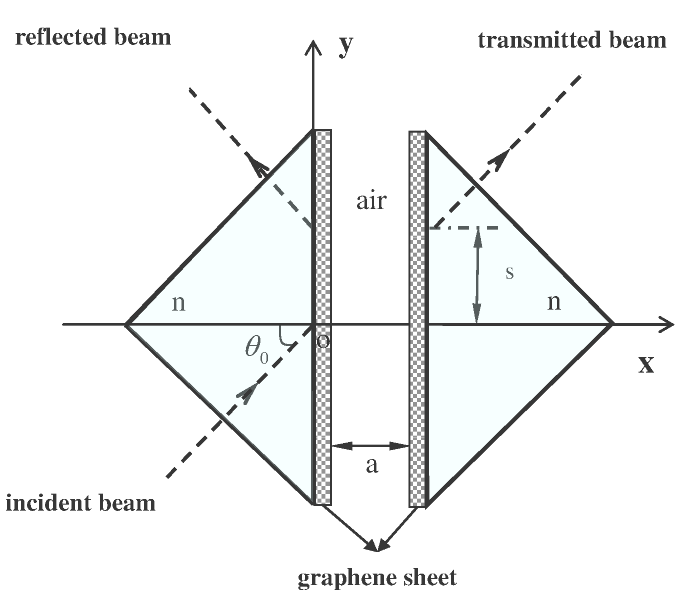}}
\caption{(Color online) Schematic diagram of GH shift in transmission in FITR configuration by coating graphene sheets, where $\theta_0$ is the central angle of incidence, $a$ is the air gap thickness and shallow parts denote the graphene sheet.}
\label{figure.1}
\end{figure}

Consider a monochromatic light beam incident upon the FIIR configuration with the central angle of incidence $\theta_0$, see Fig. \ref{figure.1}, where two dielectric prisms denoted by refractive index $n$ are separated by air gap ($n_0$) with the width $a$, and two single sheets of graphene are respectively coated to each prism at $x=0$ and $a$. In a two-dimensional case, the plane wave component of TM-polarized incident beam is assumed to be $H_{in}=\exp[i( k_{x} x + k_{y} y)]\hat{z}$,
where the wave vector $\vec{k}=(k_{x}, k_{y})=(k \cos\theta, k \sin\theta)$ with $k =n k'$, and $\theta$ is the incidence angle of plane wave component under consideration. Consequently, the reflected and transmitted waves can be expressed as
$ H_{r}=r \exp[i( - k_{x} x + k_{y} y)]\hat{z}$, and
$H_{t}= t \exp[i( k_{x} (x-a) + k_{y} y)]\hat{z}$,
with reflection and transmission coefficients $r$ and $t$. 
When the incidence angle $\theta_0 < \theta_c = \arcsin (n_0/n)$, the propagating wave in the air gap region is 
$H=[f \exp(ik'_{x} (x-a))+ g \exp(-ik'_{x}(x-a))]\exp(ik'_{y}y)\hat{z}$, where $f$ and $g$ are amplitudes of the forward and backward waves inside the air gap, the corresponding wave vector $\vec{k'}=(k'_{x},k'_{y})=(k' \cos\theta', k'\sin\theta')$, $\theta'$ is determined by $n \sin \theta =n_0 \sin \theta'$, and $k' = 2 n_0 \pi/\lambda = n_0 \omega/c$ with the wavelength $\lambda$ (frequency $\omega$) of incidence light beam and the speed of light, $c$, in vacuum. On the other hand, the electric fields in different regions are calculated from $\bigtriangledown \times \vec{H}=-i\omega \epsilon_{1(0)} \vec{E}$, where
$\epsilon_{1(0)}$ is the permittivity of dielectric prism (air gap).   According to the boundary conditions 
$\hat{n} \times(E_{2}-E_{1})=0$, and $\hat{n} \times(H_{2}-H_{1}) =J$ (surface current density $J=\sigma E$) at the interface $x=0$ and $a$ \cite{Zhan}, the coefficient $t$ is thus obtained as
\begin{equation}
\label{transmission}
t=\frac{1}{M \cos{k'_{x}a}- i N\sin{k'_{x}a}},
\end{equation}
where $ M=1+\sigma\eta_{1}$, $ N=(1/2) (\eta_{2}/\eta_{1}+\eta_{1}/\eta_{2}+2\sigma\eta_{2}+\sigma^{2}\eta_{1}\eta_{2})$, $\eta_{1}=k_{x}/\omega\epsilon_{1}$, and $\eta_{2}=k'_{x}/\omega\epsilon_{0}$. Noting that when $\sigma =0$, 
the expression (\ref{transmission}) trends to the case of ordinary FITR configuration. 
With these results, the transmissivity is finally obtained as $T= |t|^2$, and the GH shift for a well-collimated beam can be calculated, by the stationary phase approach \cite{Artmann}, as \cite{Steinberg-C}
\begin{equation}
\label{definition}
s=\mbox{Im}\left(\frac{\partial \ln t}{\partial k_y}\right)|_{\theta=\theta_{0}}.
\end{equation}
All expressions are valid by replacing by $k'_{x} = i \kappa$, 
when the incidence angle $\theta_0 > \theta_c$.
In what follows we will first calculate the resonant transmission and GH shift in FTIR configuration 
coated by graphene sheet, and then compare with the case of metal quantum well, when $\theta_0 > \theta_c$.


\begin{figure}[tbp]
\centering
\fbox{\includegraphics[width=\linewidth,height=8cm]{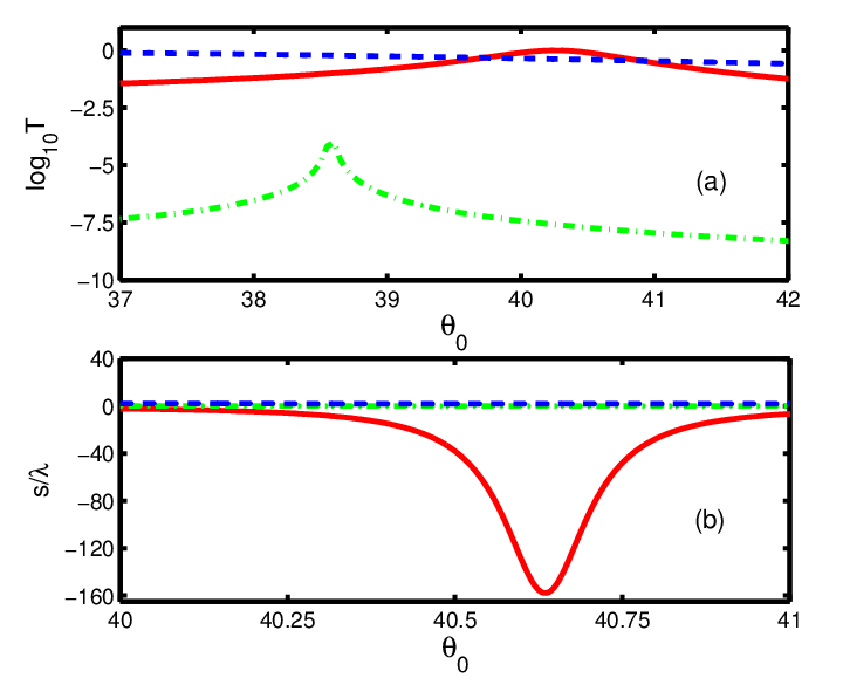}}
\caption{(Color online) Transmittivity (a) and GH shift (b) versus the incidence angle $\theta_{0}$ in FTIR configuration coating with graphene (solid red) and metal quantum well (dotted-dash green) in the THz frequency region. The ordinary FTIR configuration (dashed blue) is also compared. Parameters: $n$ = 1.605, $n_{0}$ = 1, $a = 2.5\times10^{-4}$ m, $\lambda = 3\times10^{-4}$ m, $d$ = 5 nm, $\mu$ = 0.7 eV, $\sigma = 0.021i$ S/m  (graphene), and $\epsilon_{m} = -3.62\times10^{-5} + 2.05\times10^{-5}i$ (bulk silver).}
\label{figure.2}
\end{figure}

In the THz frequency region, the optical conductivity of monolayer graphene coincides with the form of intraband electron-phonon scattering process \cite{Hanson}
\begin{equation}
\sigma = \sigma_{intra} =i\frac{e^{2}k_{B}T}{\pi\hbar^{2}(\omega+i\tau^{-1})}\left[\frac{\mu}{k_{B}T}+2 \mbox{ln}(e^{-\mu/k_{B}T}+1)\right],
\end{equation}
where $\mu$ is the chemical potential, $k_{B}$ is the Boltzmann constant, $T$ is the temperature, and $\tau$ is the electron relaxation time. In the case of high-doped graphene, $\mu\gg k_{B}T$, at least $500$ K, for all temperatures of interest, the conductivity is \begin{equation}
\label{conductivity}
\sigma(\omega)=\frac{ie^2\mu}{\pi\hbar(\omega+i\tau^{-1})},
\end{equation}
which gives the expression of the equivalent dielectric constant as follows
\beq 
\label{dielectric constant}
\epsilon = 1+i\sigma (\omega) /\omega\epsilon_{0}t_{g},
\eeq
with the effective thickness, $t_g$ ($t_g \approx 0.5$ nm). 
Obviously, when $\omega\gg\tau^{-1}$,  the dielectric constant mentioned above becomes real, 
so the loss is negligible. Otherwise, the energy dissipation is  pronounced, 
arising from considerable electron relaxation process.

Figure \ref{figure.2} (a) demonstrates that transmittivity can be significantly enhanced by graphene sheet in THz frequency region, as compared to the ordinary FIIR configuration. 
In usual case of FTIR configuration, the transmittivity decays exponentially with increasing incidence angle or thickness of air gap. However, when the graphene is coated upon the prisms,
the resonant tunneling exists in FITR, and the transmittivity can be increased up to $1$ at a certain incidence angle regardless of loss. To understand it, one can write down the reflectivity $R=1-T$, and the condition for $R=0$ gives the following dispersion relation:
\begin{equation}
\label{condition}
\sigma\eta_{1}\cosh{(\kappa a)}-\frac{1}{2}\left(\frac{\eta_{2}}{\eta_{1}}-\frac{\eta_{1}}{\eta_{2}}-\sigma^{2}\eta_{1}\eta_{2}\right)\sinh{(\kappa a)}= 0,
\end{equation}
which is different from that in graphene waveguide \cite{Hanson}. Moreover, it can be directly reduced to the condition for graphene-dielectric interface \cite{Bludov},
$\sigma\eta_{1}-(1/2)( \eta_{2}/\eta_{1}- \eta_{1}/\eta_{2}-\sigma^{2}\eta_{1}\eta_{2})= 0$, when $\kappa_{0}a \rightarrow\infty$.
When the condition (\ref{condition}) is satisfied, one mode of surface plasmon due to graphene sheet will be excited, resulting in resonant tunneling with $T=1$. 

\begin{figure}[tbp]
\centering
\fbox{\includegraphics[width=\linewidth,height=8cm]{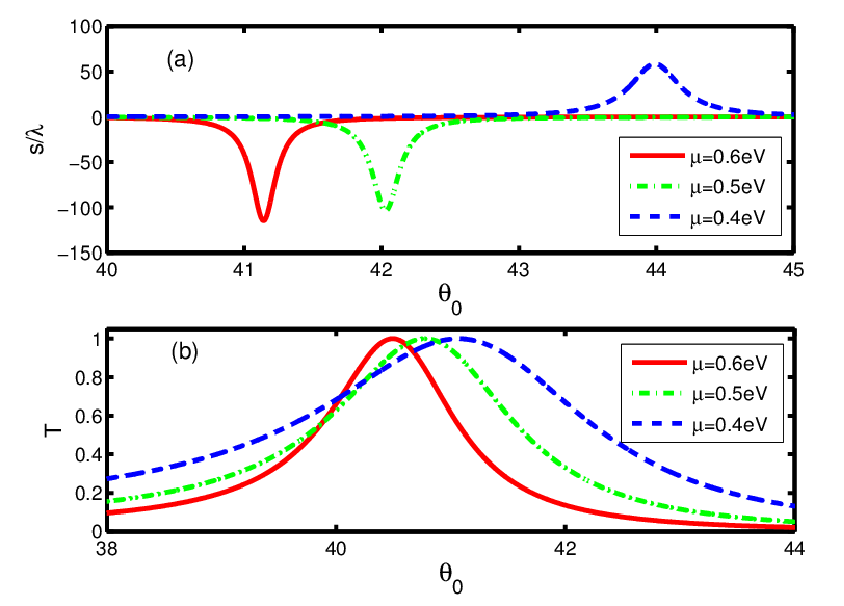}}
\caption{(Color online) GH shift (a) and transmittivity (b) versus the incidence angle $\theta_0$, where $\mu$ = 0.6 eV (solid red), $\mu$ = 0.5 eV (dot-dashed green) and $\mu$ = 0.4 eV (dotted blue). Other parameters are the same as those in Fig. \ref{figure.2}.}
\label{figure.3}
\end{figure}

\begin{figure}[tbp]
\centering
\fbox{\includegraphics[width=\linewidth,height=8cm]{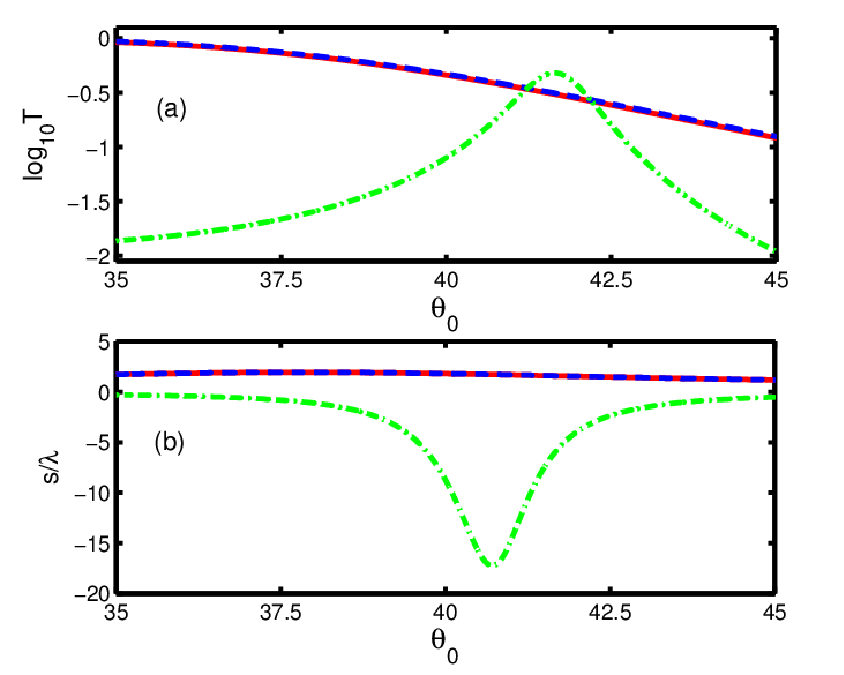}}
\caption{(Color online) Transmittivity (a) and GH shift (b) versus the incidence angle $\theta_{0}$ in FITR configuration coating with graphene (solid red) and metal quantum well (dotted-dash green) in the visible region. The ordinary FTIR configuration (dashed blue) is also compared. Parameters:
$a = 4\times10^{-7}$ m, $\lambda$ = 500 nm, $d$ = 25 nm, and $\epsilon_{m} = -2.97\times10^{-10} + 2.52\times10^{-11}i$ (bulk silver). Other parameters are the same as those in Fig. \ref{figure.2}.}
\label{figure.5}
\end{figure}

In addition, the quantum well (e.g. bulk silver) in similar FTIR configuration can enhance the transmittivity by the surface plasmon excitation as well \cite{Broe}. To see it, we consider the optical tunneling in FTIR with coating metal bulk silver) with thickness $d$. The microscopic conductivity 
is similar to Eq. (\ref{conductivity}), that is, \cite{Broe}
\beq\label{conductivity2}
\sigma=\frac{i \nu e^2}{m_e (\omega + i \tau^{-1})} \vec{U},
\eeq
where $m_e$ and $e$ are the mass and charge of electron, $\vec{U}$ is the unit tensor, and $\nu$ is the local conduction electron density at low temperature. The permittivity for silver, described by the Drude model, is approximately given in Figs. \ref{figure.2} and \ref{figure.5} for different frequencies \cite{Book}. Following \cite{Broe}, we calculate the transmittivity, shown in Fig. \ref{figure.2} (a). In principle, the surface plasmons can be excited at a certain incidence angle, so that there exists the peak of transmittivity. However, the bulk silver only works perfectly at the visible or infrared frequencies. In the region of THz frequency, the transmittivity is too small, since silver has high loss, due to the imaginary part of permittivity.

\begin{figure}[tbp]
\centering
\fbox{\includegraphics[width=\linewidth,height=8cm]{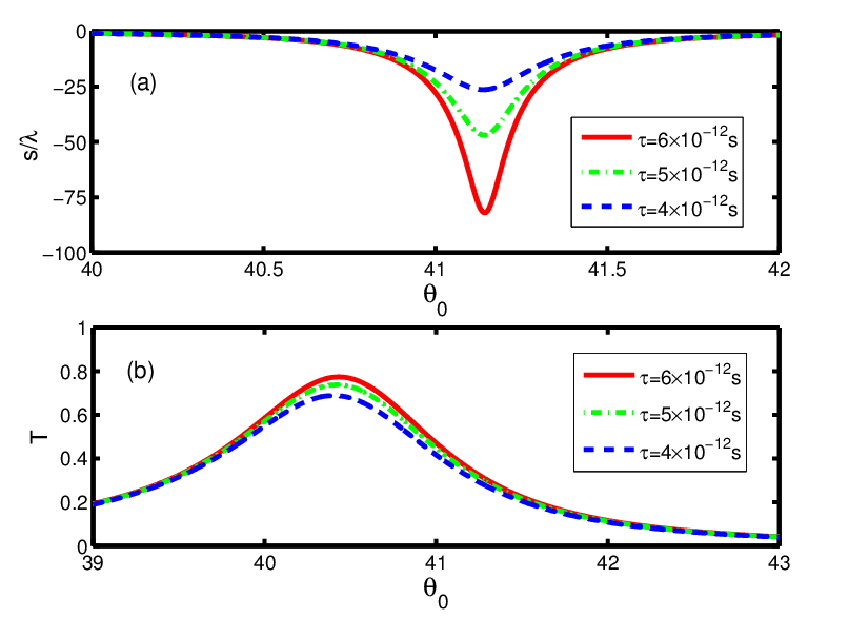}}
\caption{(Color online) GH shift (a) and transmittivity (b) versus the incidence angle $\theta_{0}$ where $\tau = 6\times10^{-12}$ s (solid red), $\tau = 5\times10^{-12}$ s (dot-dashed green) and $\tau = 4\times10^{-12}$ s (dotted blue). The chemical potential $\mu$ = 0.6 eV and other parameters are the same as those in Fig. \ref{figure.2}.}
\label{figure.4}
\end{figure}

Next, we shall turn to discuss the GH shifts. In general, GH shift (\ref{definition}) is modulated by the transmittivity $T$. But in various double-prism FTIR configurations, the behaviors are 
quite different, see Fig. \ref{figure.2} (b), where three cases are compared. 
The GH shift in ordinary FTIR configuration is the same order of magnitude as the wavelength 
due to the evanescent mode, which is relatively small. 
However, graphene sheet in FTIR configurations leads to large negative GH shifts due to the 
surface plasmon excitation and resonant tunneling. For metal (bulk silver), high loss causes the 
small absolute value of GH shifts, which is only the same order of wavelength. 
So the GH shift is indistinguishable from the case of ordinary FITR configuration. 

Depending on optical conductivity, graphene sheet has advantage over controlling GH shift by chemical potential despite dissipation \cite{JiangIEEE,Cheng}, see also Fig. \ref{figure.3} (a).
Moreover, the GH shift can be negative as well as positive with different chemical potential $\mu$.
Additionally, Fig. \ref{figure.3} (b) demonstrates that the transmittivity will be modulated correspondingly. 
Moreover, the negative and positive GH shifts can be understood by beam reshaping, resulting from the change of phase shifts for different plane components of light beam \cite{Artmann}. Besides, the GH shifts can be equivalently explained by energy flux conservation \cite{ChenOL}.

On the other hand, the GH shift and corresponding transmittivity are worthwhile to discuss 
in presence of different relaxation times. 
In Fig. \ref{figure.4}, the absolute values of peaks for GH shift and transmittivity 
become smaller when decreasing $\tau$. This is the reason that when the electronic relaxation time is small, 
the energy loss will be high. As mentioned above, when $\omega \gg \tau^{-1}$, the optical conductivity (\ref{conductivity}) turns out to be pure imaginary, thus resulting in real but negative effective dielectric constant, $\epsilon$, see Eq. (\ref{dielectric constant}). As a consequence, the maximum absolute 
values can be achieved under this circumstance. 
All the results provide the potential applications on optical spatial modulator and optical switch, e.g., by using the dependence of laterally spatial GH shifts on incidence angle. To analysis the feasibility, we calculate the angular distance $\Delta \theta_0 \approx 0.1^{\circ}$, 
the half-peak width for GH shift in Fig. \ref{figure.2} (b). 
Such value is larger than the general beam divergence $\delta \theta \approx 0.053^{\circ}$. This guarantees and determines how we can change the incidence angle. As a matter of fact, the validity of stationary phase method \cite{Artmann}, also requires $ \Delta \theta_0 \gg \delta \theta = \lambda /\pi w $ ($w$ is the local waist of beam) \cite{ChenPRE}.

Figure \ref{figure.5} further demonstrates that in the visible frequency region, the surface plasmon for bulk silver can be excited by choosing appropriate thickness $a$ \cite{Broe}. GH shift is resonantly enhanced, but not as large as the case of graphene in low THz frequency region. In addition, the resonant transmission is not perfect (less than 1), due to the fact that there still exists high loss even in such frequency domain. But graphene sheet does not support surface plasmon polaritons here. So the results are the same as those in ordinary FITR configuration.

In conclusion, we have studied that the resonance tunneling and GH shift in FTIR configuration coated by graphene sheet. The GH shift can be enhanced with perfect transmittivity due to surface plasmon excitation. As compared with metal quantum well, there are two main advantages: (i) The large GH shifts can be modulated from negative to positive values easily by changing the chemical potential varied by doping and/or an applied bias. (ii) The graphene sheet has low loss and results in large transmittivity. These overcome the difficulty in the measurement of GH shifts. All these phenomena are not only interesting for understanding the resonant tunneling in FITR configuration, the optical analogy of quantum-mechanical tunneling, but also useful for the applications in graphene-based electro-optic devices.

This work was supported by NSFC (11474193, 61176118 and 61404079), Shuguang and Yangfan (14SU35 and 14YF1408400), Research Fund for the Doctoral (2013310811003), and Eastern Scholar Programs.

\bibliography{maintext.bib}

\end{document}